\documentclass[prl,twocolumn,aps,preprintnumbers]{revtex4}
\usepackage{subfigure}
\usepackage{graphicx}
\usepackage{cancel}
\usepackage{amssymb}
\usepackage{textcomp}
\usepackage{amsmath}
\usepackage{bm}
\usepackage{times}
\usepackage{epsfig}
\usepackage{color}
\begin{document}
\preprint{CALT 68-2821}
\title{\Large Fourth Generation Bound States}
\author{Koji Ishiwata and Mark B. Wise}
\affiliation{\\  \\ California Institute of Technology, Pasadena, CA, 91125 USA}
\date{\today}
\begin{abstract}
  We investigate the spectrum and wave functions of ${\bar q}'q' $
  bound states for heavy fourth generation quarks ($q'$) that have a
  very small mixing with the three observed generations of standard
  model quarks.  Such bound states come with different color, spin and
  flavor quantum numbers.  Since the fourth generation Yukawa
  coupling, $\lambda_{q'}$, is large we include all perturbative
  corrections to the potential between the heavy quark and antiquark
  of order $\lambda_{q'}^2N_c/16\pi^2$ where $N_c$ is the number of
  colors, as well as relativistic corrections suppressed by $(v/c)^2$.
  We find that the lightest fourth generation quark masses for which a
  bound state exists for color octet states.  For the the color
  singlet states, which always have a bound state, we analyze the
  influence that the Higgs couplings have on the size and binding
  energy of the bound states.
  \end{abstract}
\maketitle
\section{I. Introduction}
Among the mysteries of nature is the number of generations. We observe
three generations, however there could be a fourth generation if the
masses of the quarks and leptons are beyond our present experimental
reach. Data from the Tevatron (under some circumstances) restricts the
masses of the $t'$ and $b'$ quarks in a fourth generation to be
greater than about $350~{\rm GeV}$. For some recent studies see
\cite{Tevatron}.  A strong constraint on the masses of fourth
generation quarks comes from precision electroweak physics. Heavy
fourth generation quarks contribute to the $S$ parameter and to the
$\rho$ parameter.  Their large contribution to the $S$ parameter rules
out a fourth generation with degenerate $t'$ and $b'$ quarks.  However
they also contribute to the $\rho$ parameter and Kribs {\it et. al.}
showed that an acceptable combined fit to precision electroweak data
can be achieved, for example, with a mass splitting of about $50~\rm
{GeV}$ between fourth generation quarks in the mass range
$350-700~{\rm GeV}$ \cite{Kribs:2007nz}.  See also the earlier work in
\cite{He:2001tp}.  Large splitting may also be possible
\cite{Holdom:1996bn}.  For a more recent discussion of electroweak
fits see \cite{Erler:2010sk,Chanowitz:2010bm,Eberhardt:2010bm}.

In addition, there is a ``unitarity upper limit'' on fourth generation
quark mass of about $ 500~{\rm GeV}$
\cite{Chanowitz:1978uj,Chanowitz:1978mv}. This does not, however,
necessarily forbid heavier quark masses. Rather it indicates that
higher order perturbative corrections become important at this mass
\cite{Chanowitz:1995rv}. Dynamical considerations rather than
unitarity give an upper bound of about $ 3~{\rm TeV}$.  This is
similar to the upper bound on the Higgs scalar mass
\cite{Dashen:1983ts,Einhorn:1986za}.

A heavy fourth generation can destabilize electroweak symmetry
breaking (see \cite{Frampton:1999xi} for a review). According to the
recent work in \cite{Hashimoto:2010at}, if there is no new physics
(apart from the fourth generation) below a ${\rm TeV}$, the Higgs mass
should be roughly equal to or larger than fourth generation quark mass
in order to avoid the instability.  Of course there could be new
particles beyond the fourth generation fermions below a ${\rm TeV}$
that get a large part of their masses from electroweak symmetry
breaking. For example, scalars $S$ that have a term in the scalar
potential $gS^{\dagger}SH^{\dagger}H$ get a contribution to the
squares of their masses equal to $g v^2/2$ $(v\simeq 246~{\rm GeV})$
and such interactions could help stabilize the Higgs potential.

It is easy to imagine simple physical mechanisms that suppress the
mass mixing between the heavy fourth generation quarks and the three
generations of standard model quarks. For example, the fourth
generation quarks and leptons could have a different value for $B-L$
than the standard three generations. (Here $B$ and $L$ are baryon
number and lepton number.) If $B-L$ violation is small then the mixing
between fourth generation quarks and the standard three generations is
suppressed.  For example, a fourth generation of quarks and leptons
with both baryon and lepton number minus three times those of the
ordinary three generations of quarks and leptons can cancel the baryon   
and lepton number anomalies allowing those symmetries to be gauged
\cite{FileviezPerez:2010gw}.

Heavy fourth generation quarks feel a strong attractive force from
Higgs exchange in both the ${\bar q}' q'$ and $q' q'$ channels that
gives rise to bound states \cite{Hung:2009hy}.  If the fourth
generation quarks have a very small mixing with the ordinary quarks,
they can be long enough lived that bound ${\bar q}' q'$ states decay
through ${\bar q'} q'$ annihilation and not via $q'$ decay to a lower
generation quark and a $W$ boson.  In this case the production of
these bound states at the LHC may have important experimental
consequences. Furthermore the $q'q'$ bound states may be long very
lived. References
\cite{Hung:2009hy,Hung:2009ia,Holdom:2009rf,Holdom:2010za} discuss
some other interesting possible physical consequences of a heavy
fourth generation.

In this paper we focus on the physics of the $\bar q' q'$ states.
Here we explore the role of perturbative corrections suppressed by
$N_c \lambda_{q'}^2/16\pi^2$, and $\alpha_s$ (here $N_c$ is the number
of color and $\alpha_s$ is strong coupling constant), as well as
relativistic corrections on the the spectrum and wave functions of the
${\bar q}' q'$ bound states. We find that the perturbative and
relativistic corrections have a significant impact on wave functions
and spectrum of ${\bar q}' q'$ bound states.

The ${\bar q}' q'$ bound states can be in a color singlet or color
octet configuration. For the color octet states we find the lightest
fourth generation quark masses for which a bound state exists. In any
color singlet configuration there is always a bound state. Therefore
we discuss the impact of the Higgs couplings to the heavy quarks on
the shape of the wave functions for the bound states and the bound
state binding energies.  In the numerical analysis, we sometimes show
results for values of $m_{q'}$ that are below the experimental limit
of $350~{\rm GeV}$ or above $500~{\rm GeV}$ where we expect
perturbation theory to be of limited use. Our excursion in to these
regimes is for pedagogical reasons and does not mean we dismiss the
constraints from experiment or the limitations imposed by
perturbativity.

\section{II. Hamiltonian}

Since precision electroweak physics favors a small value for
$|(m_{b'}-m_{t'})/(m_{b'}+m_{t'})|$ (here $m_{t'}$ and $m_{b'}$ are
masses of $t'$ and $b'$, respectively), we work in the limit where the
heavy fourth generation quark masses are equal, {\it i.e.},
$m_{t'}=m_{b'} =m_{q'}$. It is straightforward to add in the effects
of the difference between the heavy fourth generation quark masses.
(We discuss the impact of a fourth generation quark mass splitting at
the end of this paper.)  Then the leading order Hamiltonian for heavy
quark bound states from Higgs scalar exchange is
\begin{equation}
H^{(0)}={{\bf p}^2 \over m_{q'}}-
\left(\sqrt{2}G_F m_{q'}^2\right){e^{-m_hr}\over 4 \pi r},
\end{equation}
where ${\bf p}$ and ${\bf x}~(r=|{\bf x}|)$ are momentum and relative
coordinate in the center of mass frame, and $m_h$ is the Higgs scalar
mass.  In momentum space the leading potential from tree-level Higgs
exchange is
\begin{equation}
{\tilde V}({\bf p})
=-{\sqrt{2}G_F m_{q'}^2  \over {\bf p}^2 +m_h^2 }.
\end{equation}
Here we have expressed the heavy quark Yukawa coupling,
$\lambda_{q'}$, in terms of the Fermi constant $G_F$ and the heavy
fourth generation quark mass.  Using the variational method based on a
trial wave function $\Psi \propto e^{-r/a}$ \cite{Hung:2009hy} and
taking, for example, $m_h=130~{\rm GeV}$, this Hamiltonian has an
$S$-wave bound state for for $m_{q'} > 583 ~{\rm GeV}$. In this
section, we compute perturbative corrections and relativistic
corrections to the Hamiltonian.  We also give the QCD potential at
order of $\alpha_s$ for color singlet and color octet configurations.

\subsection{A. Perturbative Corrections to the Potential \\
Enhanced by the Number of Colors}

\begin{figure}[t]
  \begin{center}
    \includegraphics[scale=0.5]{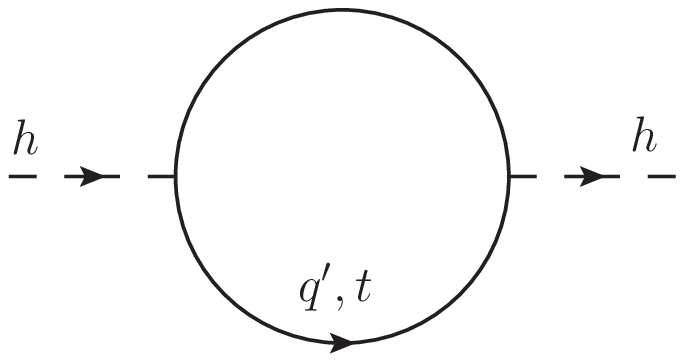}
    \includegraphics[scale=0.5]{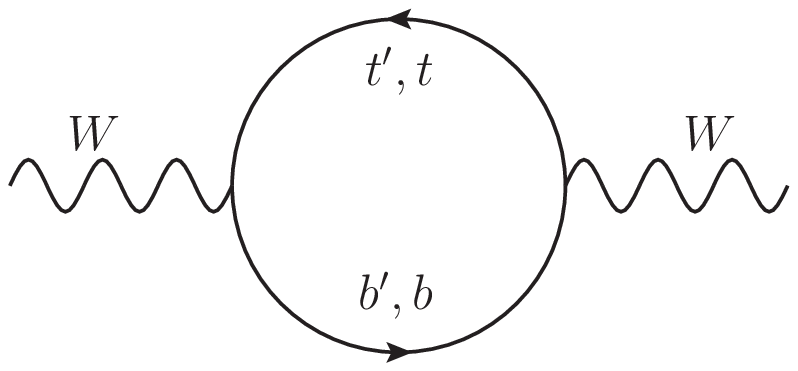}
  \end{center}
  \caption{Diagrams of Higgs self energy (left) and $W$ boson vacuum
    polarization (right).}
  \label{fig:selfegy}
\end{figure}

Here we include perturbative corrections to the potential of
$\bar{q}'q'$ state of order $N_c \lambda_{q'}^2/16 \pi^2$. These arise
from the heavy fourth generation quark contribution to the Higgs boson
self-energy $\Sigma^h(p^2)$. They will give a correction to the
leading order potential.  For pedagogical reasons we also include
terms proportional to the top quark Yukawa squared but set the other
quark masses to zero. (In the numerical results we will see the top
quark contribution is negligible.)  Expressing the Yukawa coupling
squared in terms of the quark mass squared and $G_F$, all the
perturbative corrections enhanced by a factor of $N_c$ come from the
Higgs scalar self-energy and the $W$ boson vacuum polarization.

Let us consider the Higgs propagator. It is determined by the one-loop
calculation of quark loop diagram (see left on
Fig.~\ref{fig:selfegy}),
\begin{eqnarray}
 D_h(p^2)=\frac{i(1+\delta^{h})}{p^2-m_{h}^2 -\overline{\Sigma}^{h}(p^2)},
\end{eqnarray}
with
\begin{eqnarray}
\label{selfdenom}
&&\overline{\Sigma}^{h}(p^2)
= \sum_{q=t^{\prime},b^{\prime},t}
\frac{\lambda_q^2N_c}{16\pi^2}
\left[
L^q(p^2)-L^q_<(m_h^2) \right. \nonumber \\
&&\left.  -(p^2-m_h^2) \frac{dL^q_<(p^2)}{dp^2}\Big|_{p^2=m_h^2}
\right].
\end{eqnarray}
\begin{widetext}
Here the functions $L^q$ and $L^q_{<}$ are defined by
\begin{eqnarray}
L^q(p^2) = \left\{ \begin{array}{lll}
L^q_>(p^2) & =p^2 \beta^3 
\Bigl[ \log \Bigl(\frac{1+\beta}{1-\beta} \Bigr) - i \pi \Bigr]
 &4m^2_q<p^2   \\[2mm]
L^q_<(p^2) & = -2p^2 b^3  
\tan^{-1} \Bigl(\frac{1}{b} \Bigr) &0<p^2<4m^2_q   \\[2mm]
L^q_{-}(p^2) & = p^2\beta^3  
     \log \left(\frac{\beta+1}{\beta-1} \right) &p^2<0
\end{array} \right. ,
\end{eqnarray}
where
\begin{equation}
\beta= \sqrt{1-{4m_q^2 \over p^2}}, ~~ 
b=\sqrt{{4 m_q^2\over p^2}-1}~~{\rm and}~~x_q= {m^2_q \over m^2_h}.
\end{equation}
The derivative of $ L^q_<(p^2)$ evaluated at $p^2=m_h^2 $ is
\begin{eqnarray}
\frac{dL^q_<(p^2)}{dp^2}\Big|_{p^2=m^2_h} &=&
1-4x_q
+2\sqrt{4x_q-1}(1+2x_q)\tan^{-1}(1/\sqrt{4x_q-1}).
\end{eqnarray}
Finally, using ${\overline{\rm MS}}$ subtraction (with the number of
space-time dimensions $n=4-\epsilon$), $\delta^{h}$ is given as
\begin{equation}
\delta^{h} = \frac{d\Sigma^{h}(p^2)}{dp^2}|_{p^2=m_h^2}=
\sum_{q=t^{\prime},b^{\prime},t} \frac{\lambda_q^2N_c}{16 \pi^2}
\biggl[
-\Bigl(\frac{2}{\epsilon}+\log\Bigl(\frac{\mu^2}{m_q^2}
\Bigr)\Bigr)
-2+ \frac{dL^q_<(p^2)}{dp^2}|_{p^2=m^2_h}\biggr].
\end{equation}
\end{widetext}
Expanding the factor $\overline{\Sigma}^{h}(p^2)$ in a power series in,
$p^2-m_h^2$, it is clear from its definition in Eq.~(\ref{selfdenom})
that it first contributes at order $(p^2-m_h^2)^2$.

In the numerator of the Higgs propagator the factor $\delta^h$ is
divergent.  This divergence is cancelled in the potential if we
express the fourth generation quark Yukawa couplings
$\lambda_{q'}^{2}$ in terms of $G_F m_{q'}^2$.  The resulting
correction to the Fourier transform of the potential is
\begin{equation}
\label{perpot1}
{\tilde V}_{\rm pert}({\bf p})=-{\sqrt{2} G_Fm^2_{q'}
\delta \over {\bf p^2}+m_h^2}.
\end{equation}
Here the perturbative corrections in the denominator of the Higgs
propagator can be negligible. This arises because of a cancellation
between the three terms in Eq.~(\ref{selfdenom}).  Here we have used
the expansions,
\begin{eqnarray}
&&L_<^q(m_h^2) =m_q^2\left(-8+{8 \over 3}{m_h^2 \over m_q^2} +\ldots \right),
\label{eq:L<}
\\ &&
\frac{dL^q_<(p^2)}{dp^2}\Big|_{p^2=m^2_h}=-{8 \over 3}+\ldots ,
\\ &&
L_-^q(-{\bf p}^2)=
m_q^2\left(-8 -{8 \over 3}{{\bf p}^2 \over m_q^2}+\ldots \right).
\label{eq:L_}
\end{eqnarray}
Even though expansions we used in Eqs.~(\ref{eq:L<})-(\ref{eq:L_}) are
applicable for $m_h\ll 2m_{q'}$, we have checked numerically that the
denominator of the Eq.~(\ref{perpot1}) is a good approximation when
$m_h\sim m_{q'}$.  On the other hand, the factor $\delta$ in the
numerator of the potential is given by
\begin{equation}
\label{delta}
\delta=\delta ^h+{{\Pi}^T_{WW}(0) \over M_W^2}.
\end{equation}
Here $\Pi^T_{WW}(p^2)$ is the transverse part of $W$ boson vacuum
polarization $\Pi_{WW,\mu\nu}(p^2)$, defined by
$\Pi_{WW,\mu\nu}(p^2)=g_{\mu\nu}\Pi^T_{WW}(p^2)+\cdots$. For
${\Pi}^T_{WW}(0)$ there are two contributions (see right on
Fig.~\ref{fig:selfegy}),
${\Pi}^T_{WW}(0)={\Pi}^T_{WWq'}(0)+{\Pi}^T_{WWt}(0)$ and these are
\begin{eqnarray}
&&{\Pi}^T_{WWq'}(0)/M^2_W={2 \lambda_{q'}^2 N_c \over 16 \pi^2}
\left( {2 \over \epsilon} +{\rm ln} \left({\mu^2 \over m_{q'}^2} 
\right) \right),
\\ &&
{\Pi}^T_{WWt}(0)/M^2_W
={ \lambda_{t}^2 N_c \over 16 \pi^2}\left( {2 \over \epsilon} 
+{\rm ln} \left({\mu^2 \over m_{t}^2} \right) +{1 \over 2} \right).
\end{eqnarray}
Using the above results, one obtains
\begin{eqnarray}
\label{delta}
&&\delta={2N_c \over 48 \pi^2}\sum_{q=t',b',t} \lambda_q^2
+ {N_c \over 32 \pi^2} \lambda_t^2 \nonumber \\
&&={ \sqrt{2} G_F\over 2 \pi^2} m_{q'}^2+{7\sqrt{2} G_F  
\over 16 \pi^2}m_t^2.
\end{eqnarray}
Here we take $N_c=3$. The divergences in $\delta^h$ and
${\Pi}^T_{WW}(0)$ canceled.  Eqs.~(\ref{perpot1}) and (\ref{delta})
are the main results of this section.  

\begin{figure}[t]
  \begin{center}
    \includegraphics[scale=0.5]{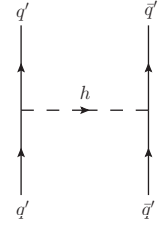}
  \end{center}
  \caption{$t$-channel Higgs exchange.}
  \label{fig:tHiggs}
\end{figure}

\subsection{B. Relativistic  Corrections }

 \begin{figure}[t]
  \begin{center}
    \includegraphics[scale=0.5]{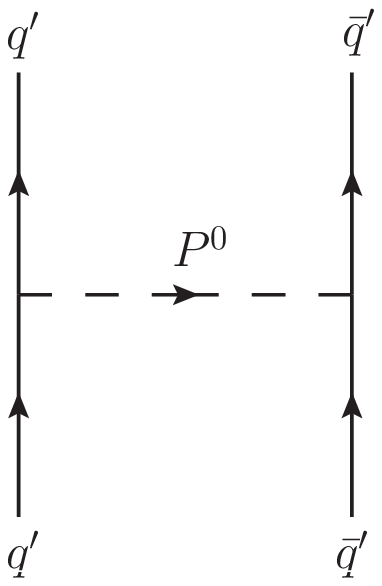}
    \includegraphics[scale=0.5]{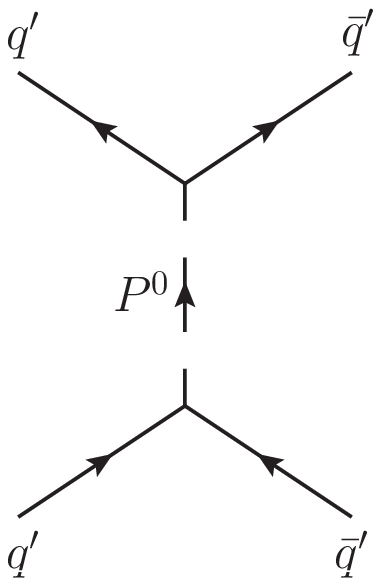}
  \end{center}
  \caption{$s$- and $t$-channel neutral fictitious scalar exchange.}
  \label{fig:nfict}
\end{figure}

\begin{figure}[t]
  \begin{center}
    \includegraphics[scale=0.5]{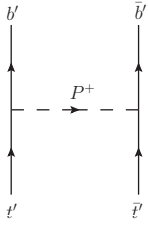}
  \end{center}
  \caption{$t$-channel charged fictitious scalar exchange.}
  \label{fig:cfict}
\end{figure}

Relativistic corrections to the potential come from expanding the
spinors in the Higgs scalar exchange diagram and from including the
contributions from longitudinal gauge bosons and the ``fictitious
scalars'' in $R_{\xi}$ gauge. We choose $\xi=1$ so that the
longitudinal gauge boson contribution vanishes. The corrections
appropriate for the ground $S$-wave bound states are given here.  For
discussions of how the relativistic corrections to the potential are
derived from Feynman diagrams see
Refs. \cite{Manohar:2000cg,Manohar:2000hj}.

Expanding the spinors for the $t$-channel Higgs exchange diagram
(shown in Fig.~\ref{fig:tHiggs}) gives the relativistic correction to
the potential,
\begin{equation}
{\tilde V}_{\rm rel. Higgs}({\bf p})
=-{\sqrt{2}G_F \over 4}\left( {{\bf p}^2 \over {\bf p}^2+m_h^2} \right).
\end{equation}

Neutral fourth generation bound states can exist in the flavor states
${\bar t}'t'$ and ${\bar b}'b'$, can be in color singles (${\bf 1}$)
and octets (${\bf 8}$), and furthermore they can have zero and one
spins. Since we are working in the limit $m_{t'}=m_{b'}$, it is
convenient to decompose the flavor structure into heavy quark isospin
$I=0$ ({\it i.e.}, $({\bar t'}t'+{\bar b'}b')/\sqrt{2}$) and $I=1$
({\it i.e.}, $({\bar t'}t'-{\bar b'}b')/\sqrt{2}$).  So far the
contributions to the Hamiltonian have not depended on the bound states
heavy quark isospin, color and spin quantum numbers.  However, the
contributions we consider now do depend on these quantum numbers. We
therefore attach the superscript, $(C={\rm color},I={\rm heavy~ quark~
  isospin},S={\rm heavy ~quark~ spin})$ to the potential. Since these
eight states characterized by ($C,~I,~S$) do not necessarily form
bound states, hereafter we call them ``channels''.

\begin{figure}[t]
  \begin{center}
    \includegraphics[scale=0.5]{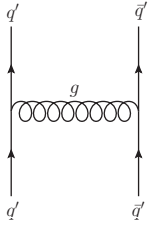}
  \end{center}
  \caption{Gluon exchange.}
  \label{fig:gluon}
\end{figure}

Exchange of the neutral `` fictitious scalar'' (which we call $P^0$)
in the $t$-channel (left on Fig.~\ref{fig:nfict}) gives spin-dependent
potential, but is independent of the color and flavor. We find that,
\begin{equation}
 {\tilde V}^{(S)}_{P^{0},t\text{-channel}}({\bf p})
={\sqrt{2}G_F \over 4}{{\bf p}^2 \over {\bf p}^2+M_Z^2}\Omega(S),
\end{equation}
where $M_Z$ is $Z$ boson mass, and for spin one, $\Omega(1)=-1/3$, and
for spin zero, $\Omega(0)=1$. This contribution is attractive in the
spin one channel and repulsive in the spin zero channel.  The
$s$-channel $P^0$ exchange (left on Fig.~\ref{fig:nfict}), on the
other hand, gives a repulsive potential which only occurs in the
$({\bf 1},1,0)$ channel,
 \begin{equation}
 {\tilde V}^{({\bf 1},1,0)}_{P^{0}, s\text{-channel}}({\bf p})
={3\sqrt{2}  G_F\over 1-M_Z^2/(4 m_{q'}^2)}.
\end{equation}
It is enhanced by a factor of $N_c=3$, compared with other
relativistic corrections to the potential.

The final relativistic correction to the potential comes from
$t$-channel exchange of the charged fictitious scalar $P^+$, which is
depicted in Fig.~\ref{fig:cfict}. It is independent of color but
depends on spin and on flavor since it mixes the ${\bar t}'t'$ and
${\bar b}'b'$ channel. We find that,
\begin{eqnarray}
  {\tilde V}^{(S)}_{P^{+}}({\bf p})
=\pm{\sqrt{2}G_F \over 2}{ {\bf p}^2 \over {\bf p}^2+M_W^2}\Omega(S),
\end{eqnarray}
where $M_W$ is $W$ boson mass. Here plus and minus sign correspond to
$I=1$ and 0 channels, respectively.  

Finally there is the usual relativistic correction to the kinetic
energy,
\begin{equation}
T_{\rm rel}=-{{\bf p}^2 \over m_{q'}}\left({{\bf p}^2\over 4m^2_{q'} }\right).
\end{equation}

\subsection{C. QCD Potential}
There are also contributions to the potential from one-gluon exchange
(Fig.~\ref{fig:gluon}). They are attractive in the color singlet
channel and repulsive in the color octet channel but are spin and
flavor independent, and given as
\begin{eqnarray}
&&V^{({\bf 1})}_{\rm QCD}(r)
=-{ 4 \over 3}\alpha_s\left({1 \over  r}\right),
\\ &&
V^{({\bf 8})}_{\rm QCD}(r)
={ 1 \over 6}\alpha_s\left({1 \over r}\right).
\end{eqnarray}
There are always bound states in the color singlet channel because the
strong interactions confine.  In the octet channel there are no bound
states without the Yukawa potential from Higgs exchange.  In our
numerical work we evaluate $\alpha_s$ at the $Z$ boson mass,
$\alpha_s(M_Z)=0.118$.  

\section{III. Numerical Results}
In this section we discuss the ground state $S$-wave states in the
various color, heavy flavor isospin, and spin channels. The
Hamiltonian for this system is
\begin{equation}
H=H^{(0)}+H^{(1)},
\end{equation}
where
\begin{eqnarray}
H^{(1)}&=&T_{\rm rel}+V_{\rm pert}+V_{\rm rel. Higgs}+V_{P^{0},t\text{-channel}}
\nonumber \\ &&
 +V_{P^{0},s\text{-channel}}+V_{P^{+}}+V_{\rm QCD}.
\end{eqnarray}
We use the variational method, minimizing $E[a]=\langle \psi |H| \psi
\rangle /\langle \psi | \psi \rangle$ for trial wave functions $\psi
\propto e^{-r/a}$.  In order for the $v/c$ expansion to make sense, we
restrict our analysis to wave functions that give an expectation value
for ${\bf p}^2/m_{q'}^2$ that is smaller than 1/3. This ensures that
higher order terms in the $v/c$ expansion, which we have neglected,
are not important. This means that,
\begin{eqnarray}
a^2 \ge 3/m_{q'}^2.
\label{eq:xregion}
\end{eqnarray} 

Before discussing the numerical results, we give the formula for
$E[a]$ in each channel. The expectation values of the kinetic energy
and the potential from the Higgs exchange and $t$-channel neutral
fictitious scalar exchange give a common contribution for color
singlet/octet and isospin zero/one channels. These are given by,
\begin{widetext}
\begin{eqnarray}
E^{(S)}_{\rm com}[a]=
{1\over  a} \left[ {1 \over m_{q'}a}-{5 \over 4 m_{q'}^3 a^3} \right]
-{\sqrt{2}G_Fm_{q'}^2 \over \pi a}
\left[
{1+\delta-m_h^2/4m_{q'}^2\over (2+am_h)^2}
+{1 \over  4 m^2_{q'}a^2}
-{\Omega(S)\over 4}\left\{
{1 \over   m^2_{q'}a^2}-
{ M_Z^2/m_{q'}^2 \over (2+aM_Z)^2}
\right\}
\right].
\end{eqnarray}
The first term comes from the kinetic energy, while first and second
terms in the second parentheses are from $t$-channel Higgs exchange,
including the perturbative correction to the Higgs propagator. The
rest is from neutral fictitious scalar exchange in $t$-channel.  The
$s$-channel neutral fictitious scalar exchange, on the other hand,
gives a contribution only for the $({\bf 1},1,0)$ channel, which is
\begin{eqnarray}
E^{({\bf 1},1,0)}_{P^0,s\text{-channel}}[a]={\sqrt{2}G_Fm_{q'}^2 \over \pi a}
{3 \over (1- M_Z^2/4m_{q'}^2)}{1 \over m_{q'}^2 a^2}.
\end{eqnarray}
As we mentioned, this term always contributes as a positive
(repulsive) term in total energy, and it is enhanced by color factor
$N_c=3$. Charged fictitious scalar exchange gives
\begin{eqnarray}
E^{(S)}_{P^+}[a]={\sqrt{2}G_Fm_{q'}^2 \over \pi a}
\Omega(S)\left[
{1 \over 2 m_{q'}^2 a^2}- {M_{W}^2/2m^2_{q'} \over (2+aM_W)^2}
\right].
\end{eqnarray}
Finally, the contribution from one gluon exchange is
\begin{eqnarray}
E^{(\bf 1)}_{\rm QCD}[a]=-{4\alpha_s \over 3 a},\ \ 
E^{(\bf 8)}_{\rm QCD}[a]={\alpha_s \over 6 a}.
\end{eqnarray}
\end{widetext}
With all terms we have given above, the variational energy in each 
channel is obtained,
\begin{eqnarray}
E^{({\bf 1},0,S)}[a]&=&E^{(S)}_{\rm com}+E^{(S)}_{P^+}+E^{(\bf 1)}_{\rm QCD},
\\
E^{({\bf 1},1,0)}[a]&=&E^{(0)}_{\rm com}-E^{(0)}_{P^+}+E^{({\bf 1},1,0)}_{P^0,s}
+E^{(\bf 1)}_{\rm QCD},
\\
E^{({\bf 1},1,1)}[a]&=&E^{(1)}_{\rm com}-E^{(1)}_{P^+}+E^{(\bf 1)}_{\rm QCD},
\end{eqnarray}
for color singlet state, $({\bf 1},I,S)$, and
\begin{eqnarray}
E^{({\bf 8},0,S)}[a]&=&E^{(S)}_{\rm com}+E^{(S)}_{P^+}+E^{(\bf 8)}_{\rm QCD},
\\
E^{({\bf 8},1,S)}[a]&=&E^{(S)}_{\rm com}-E^{(S)}_{P^+}+E^{(\bf 8)}_{\rm QCD},
\end{eqnarray}
for color octet state, $({\bf 8},I,S)$.  These results are summarized
in the Appendix.

\begin{figure}[t]
  \begin{center}
    \includegraphics[scale=1.]{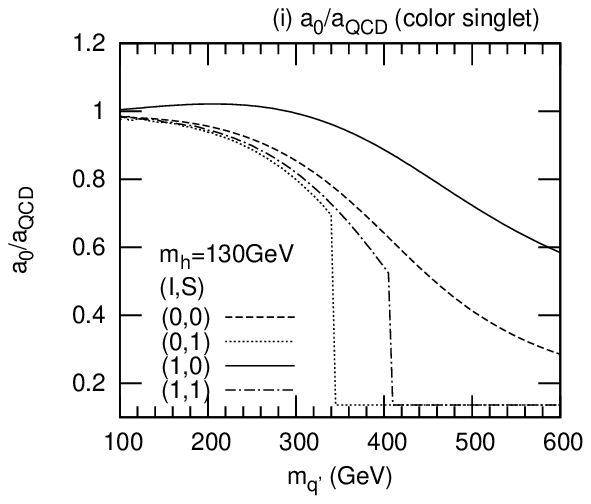}
    \includegraphics[scale=1.]{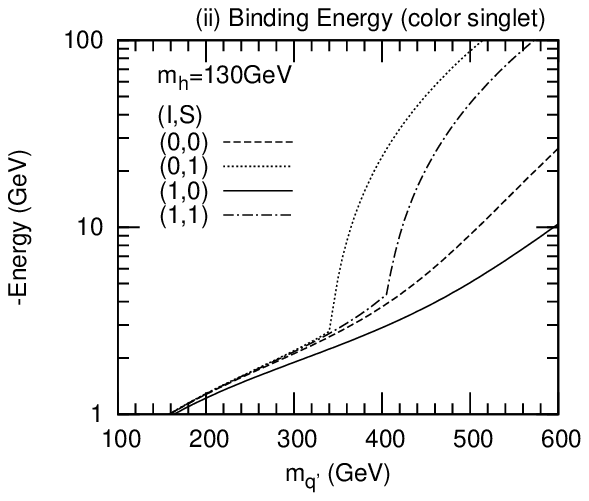}
  \end{center}
  \caption{(i) $a_0/a_{\rm QCD}$ as the function of quark mass in
    color singlet channels. Here we take $m_h=130~{\rm GeV}$ and $a_0$
    is the value for which $E[a]$ is minimized (and negative) for
    fixed quark mass. (ii) Variational binding energy of color singlet
    channels as the function of the heavy quark mass. Here we set
    $a=a_0$. }
  \label{fig:C1}
\end{figure}

We compute the variational energy $E^{(C,I,S)}[a]$ in each channel and
study the properties of the bound states. In the color singlet
channels, there always exists a bound state.  For small enough
$m_{q'}$ the state is very close to the familiar QCD ``onium"
states. However as $m_{q'}$ increases the parts of the potential
proportional to $m_{q'}^2$ become more important.  We find the value
of $a$ (in the parameter region given by Eq.~(\ref{eq:xregion})) which
gives minimum binding energy for fixed $m_{q'}$. (We denote call it
$a_0$.)  It is compared with Bohr radius of pure QCD potential,
$a_{\rm QCD}\equiv 2 \alpha_s/3m_{q'}$. We begin by taking
$m_h=130~{\rm GeV}$.  Later we redo the analysis for the case
$m_h=m_{q'}$ which may provide more realistic values of the Higgs mass
given the constraints from stability of the Higgs potential.  The
results (for $m_h=130{\rm GeV}$) are shown in the upper panel of
Fig.~\ref{fig:C1}.  From this figure, it can be seen that the size of
the singlet bound states are not close to a QCD-like bound state when
$m_{q'}\gtrsim 400~{\rm GeV}$, except for the $({\bf 1},1,0)$
channel. The sharp break in behavior as $m_{q'}$ increases is due to
the limit we impose on the value of $a_0$ ({\it i.e.}, it is greater
than or equal to $\sqrt{ 3}/m_{q'}$), which ensures that relativistic
corrections are not too large .  In the $({\bf 1},1,0)$ channel, the
contribution from the repulsive $s$-channel $P^0$ exchange potential
is so large that the bound state has $a_0>a_{\rm QCD}$ for a range of
masses.  In the lower panel of Fig.~\ref{fig:C1}, we plot the
variational binding energy computed at $a=a_0$ for each color singlet
channel. We find binding energies of $O((10-100)~{\rm GeV})$ for
$m_{q'}\sim 400 -500~{\rm GeV}$.

\begin{figure}[t]
  \begin{center}
    \includegraphics[scale=1.]{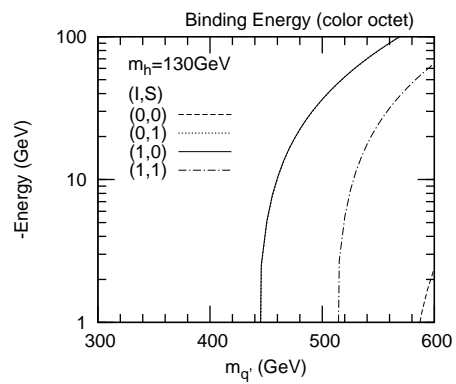}
  \end{center}
  \caption{Variational binding energy of color octet channels plotted
    as the function of the heavy quark mass. In the plot, we use
    $m_h=130~{\rm GeV}$ and take $a$ as the value which gives the
    lowest binding energy for fixed $m_{q'}$. Note that $({\bf
      8},0,1)$ and $({\bf 8},1,0)$ channels give almost the same
    results.}
  \label{fig:C8egy}
\end{figure}

For the color octet channels, on the other hand, bound states do not
exist if the heavy fourth generation quark is too light and of course
the Higgs Yukawa couplings always play a crucial role because the QCD
potential is repulsive. In our numerical analysis, we find the lowest
value of $m_{q'}$ for which the minimum of the variational energy
$E[a]$ (in the region, $a \ge \sqrt{3}/m_{q'}$) has a negative value.
The results are summarized in Table~\ref{tab:lowmass}. Note that the
values of the fourth generation quark masses relevant here are not the
ones in parenthesis.  We find that the lower limit reduces to
$440-570~{\rm GeV}$, compared to the one given by the leading order
Hamiltonian ({\it i.e.}, $583~{\rm GeV}$).  As in the color singlet
channel, we plot the lowest variational binding energy for fixed
$m_{q'}$ in Fig.~\ref{fig:C8egy}.  This figure indicates that color
octet bound states with binding energy of $O((10-100)~{\rm GeV})$
exist when $m_{q'}\simeq 450-550~{\rm GeV}$. The color octet states we
found form color singlet hadrons by neutralizing their color charge at
long distances with gluons and light quark-anti quark pairs.
\begin{table}[t]
  \begin{center}
\begin{tabular}{cc}
      \hline \hline
      \ \ \ \ ($C,~I,~S$)\ \ \ \    & Lower limit of $m_{q'}$\\
       \hline
      $({\bf 8}, 0, 0)$ & $574$ GeV ($574$ GeV) \\
      $({\bf 8}, 0, 1)$ & $440$ GeV ($359$ GeV) \\
      $({\bf 8}, 1, 0)$ & $440$ GeV ($359$ GeV) \\
      $({\bf 8}, 1, 1)$ & $510$ GeV ($439$ GeV) \\
      \hline \hline
    \end{tabular}
  \end{center}
  \caption{Lower limit of quark mass for which  a 
    bound state forms in the various color octet channels. We take 
    $m_h=130~{\rm GeV}$.The values
    in parentheses are given by using
    $a\ge \sqrt{2}/m_{q'}$ instead of Eq.~(\ref{eq:xregion}) }
  \label{tab:lowmass}
\end{table}

\begin{figure}[t]
  \begin{center}
    \includegraphics[scale=1.]{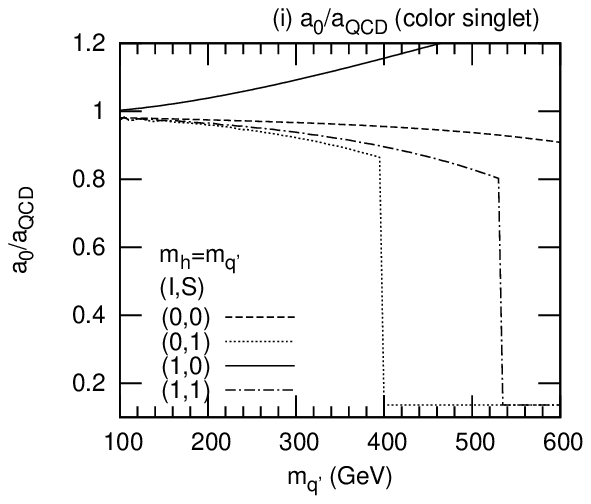}
    \includegraphics[scale=1.]{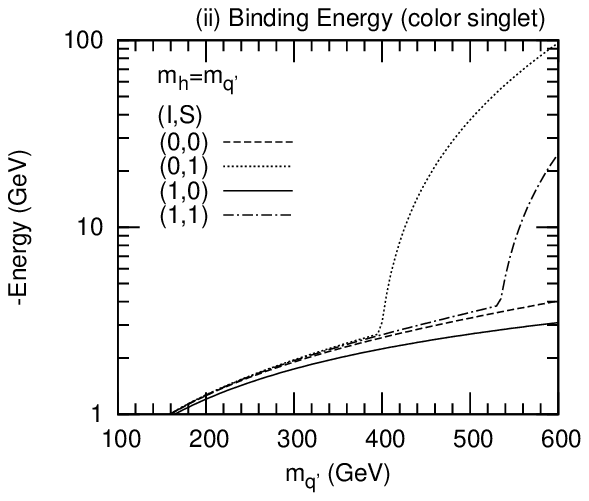}
  \end{center}
  \caption{The same as Fig.~\ref{fig:C1}, except for taking $m_h=m_{q'}$}
  \label{fig:BohrMhLr}
\end{figure}

\begin{table}[t]
  \begin{center}
\begin{tabular}{cc}
      \hline \hline
      \ \ \ \ ($C,~I,~S$)\ \ \ \    & Lower limit of $m_{q'}$\\
       \hline
      $({\bf 8}, 0, 0)$ & No bound state \\
      $({\bf 8}, 0, 1)$ & $534$ GeV \\
      $({\bf 8}, 1, 0)$ & $534$ GeV \\
      $({\bf 8}, 1, 1)$ & $696$ GeV \\
      \hline \hline
    \end{tabular}
  \end{center}
  \caption{The same as Table.~\ref{tab:lowmass}, except for taking 
    $m_h=m_{q'}$.}
  \label{tab:lowmassMhLr}
\end{table}

\begin{figure}[t]
  \begin{center}
    \includegraphics[scale=1.]{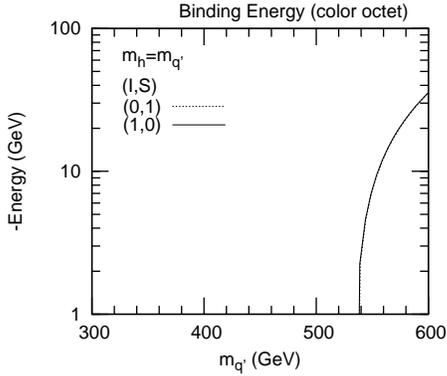}
  \end{center}
  \caption{The same as Fig.~\ref{fig:C8egy}, except for taking $m_h=m_{q'}$}
  \label{fig:C8egyMhLr}
\end{figure}

It is important to remember that when $a_0$ is at the end of the range
given by Eq.~(\ref{eq:xregion}), the actual bound state may be
relativistic and more deeply bound than the results presented in this
section indicate. Such a situation occurs for the color octet results,
except for the $({\bf 8},0,0)$ state, and in some of the color singlet
channels at larger heavy quark masses.  In order to see how our
results are affected by the choice of region for $a$, we consider, for
example, the case where expectation value of ${\bf p}^2/m_{q'}^2$ is
less than $1/2$, which corresponds to $a\ge \sqrt{2}/m_{q'}$.  In
Table~\ref{tab:lowmass}, the lower limit on $m_{q'}$ for an octet
bound state to exist is also derived using this region for $a$,
instead of Eq.~(\ref{eq:xregion}) (see the values in parentheses). As
it is shown, the limit becomes smaller by $10-20\%$ (except for the
$({\bf 8},0,0)$ state).

Finally, we give the numerical results in the case of $m_h=m_{q'}$.
In the same manner as we did when $m_h$ was fixed at $130~{\rm GeV}$,
$a_0/a_{\rm QCD}$ and the binding energy for color singlet channels
are given in Fig.~\ref{fig:C1} and the lower bound on the fourth
generation quark mass, and the binding energy for color octet channels
are given in Table~\ref{tab:lowmassMhLr} and Fig.~\ref{fig:C8egyMhLr},
respectively. For singlet states, the bound state is certainly not
QCD-like when $m_{q'}>400~(535)~{\rm GeV}$ in the $({\bf 1},0,1)$
($({\bf 1},1,1)$) channels. On the other hand, octet channels do not
form bound state unless $m_{q'}\gtrsim 535~{\rm GeV}$, which is near
the unitarity bound or equivalently strong coupling regime.

We have assumed that $m_{t'}=m_{b'}$ throughout this paper, however it
is straightforward to take into account the mass difference between
the the heavy fourth generation quarks. In that case, the $I=0$ and
$1$ sates are no longer the energy eigenstates. Rather we denote the
eigenstates by $|+\rangle$ and $|- \rangle$. They are the following
linear combinations of the heavy quark isospin eigenstates,
\begin{eqnarray}
|+ \rangle &\propto& |I=0 \rangle + B_+|I=1 \rangle ,
\\ 
|- \rangle &\propto& B_-|I=0 \rangle + |I=1 \rangle.
\end{eqnarray}
Introducing the notation, $m_{\pm}=m_{t'}\pm m_{b'}$, the energy
eigenvalues $E_{\pm}$ and mixing parameters $B_{\pm}$ are,
\begin{eqnarray}
E_{\pm}&=&m_+ + \frac{E^{(C,0,S)}+E^{(C,1,S)}}{2} 
\nonumber \\ &&
\pm\sqrt{\left( \frac{E^{(C,0,S)}-E^{(C,1,S)}}{2} \right)^2+m^2_-},
\\
B_{+}&=&\frac{m_-}{m_++E^{(C,1,S)}-E_+},
\\
B_{-}&=&\frac{m_-}{m_++E^{(C,0,S)}-E_-},
\end{eqnarray}
respectively. When $a_0\sim a_{\rm QCD}$, $m_-$ is larger in magnitude
than $(E^{(C,0,S)}-E^{(C,1,S)})/2$. (Here we are assuming $|m_ -|\sim
50~{\rm GeV}$.) Then, the mixing parameter is not neglegible. On the
other hand when $a_0$ is much smaller than $a_{\rm QCD}$, $m_{-}$ is
not important and the mixing parameter is negligible. Then the states
$| \pm \rangle$ are almost isospin eigenstates, {\it i.e.}, $|+
\rangle \simeq |I=0 \rangle$ and $|- \rangle \simeq |I=1 \rangle$. In
a more accurate evaluation, one should also take into account the
correction $m_-$ makes to $E^{(C,0,S)}$ and $E^{(C,1,S)}$. These
corrections are suppressed by, $(m_-/m_+)$ and are expected to change
the binding energies by a few to $10\%$.

\section{IV. Concluding Remarks}
Heavy fourth generation quarks may have a long enough lifetime that it
is sensible to consider their bound states.  At the LHC heavy quark
${\bar q'} q'$ bound states will be produced by gluon fusion.  Hence
it is important to understand the properties of these states.  In this
paper we have determined the binding energies and sizes of these
states.  For $m_{q'}\gtrsim 400{\rm GeV}$, the Higgs Yukawa coupling
plays a crucial role in the properties of these states and also
relativistic and perturbative corrections are important.  In a future
publication we hope to elucidate more of their properties, including
production rates at the LHC and their decay branching ratios.

\subsection*{Acknowledgment}
The work was supported in part by the U.S. Department of Energy under
contract No. DE-FG02-92ER40701, and by the Gordon and Betty Moore
Foundation.  We are grateful to B. Grinstein for useful
discussions.

\appendix
\begin{widetext}
\section{Appendix}
Here we give explicit formulas for the variational energy in each channel.
\begin{eqnarray}
E[a]^{({\bf 1},0,0)}&=&{2\sqrt{2}G_Fm_{q'}^2 \over \pi a^3}
\left[ -{(1+\delta)a^2 \over  2 (2+a m_h)^2}+{1 \over 4m_{q'} ^2}  
+ {m_h^2 \over 8 m_{q'}^2 }{   a^2 \over   (2+a m_h)^2} \right. 
\nonumber \\
&&\left .-{M_Z^2 \over 8 m_{q'}^2} {   a^2 \over   (2+a M_Z)^2}
 -{M_W^2 \over 4 m_{q'}^2} {   a^2 \over   (2+a M_W)^2}   \right]
-{4 \over 3a}\alpha_s +{1\over  a^3} \left[ {a \over m_{q'}}
-{5 \over 4 m_{q'}^3 a} \right]
\\
E[a]^{({\bf 1},0,1)}&=&{2\sqrt{2}G_Fm_{q'}^2 \over \pi a^3}
\left[ -{(1+\delta)a^2 \over   2(2+a m_h)^2}-{1 \over 4m_{q'} ^2}  
+ {m_h^2 \over 8 m_{q'}^2 }{   a^2 \over   (2+a m_h)^2} \right. 
\nonumber \\
&&\left .+{M_Z^2 \over 24 m_{q'}^2} {   a^2 \over   (2+a M_Z)^2} 
+{M_W^2 \over 12 m_{q'}^2} {   a^2 \over   (2+a M_W)^2}   \right]
-{4 \over 3a}\alpha_s +{1\over  a^3} \left[ {a \over m_{q'}}
-{5 \over 4 m_{q'}^3 a} \right]
\\
E[a]^{({\bf 1},1,0)}&=&{2\sqrt{2}G_Fm_{q'}^2 \over \pi a^3}
\left[ -{(1+\delta)a^2 \over  2 (2+a m_h)^2} -{1 \over 4m_{q'} ^2} 
+{6 \over (4m_{q'}^2-M_Z^2)} + {m_h^2 \over 8 m_{q'}^2 }
{   a^2 \over   (2+a m_h)^2} \right. 
\nonumber \\
&&\left .-{M_Z^2 \over 8 m_{q'}^2} {   a^2 \over   (2+a M_Z)^2} 
+{M_W^2 \over 4 m_{q'}^2} {   a^2 \over   (2+a M_W)^2}   \right]
-{4 \over 3a}\alpha_s+{1\over  a^3} \left[ {a \over m_{q'}}
-{5 \over 4 m_{q'}^3 a} \right]
\\
E[a]^{({\bf 1},1,1)}&=&{2\sqrt{2}G_Fm_{q'}^2 \over \pi a^3}
\left[ -{(1+\delta)a^2 \over  2 (2+a m_h)^2} -{1 \over 12m_{q'} ^2}  
+ {m_h^2 \over 8 m_{q'}^2 }{   a^2 \over   (2+a m_h)^2}\right. 
\nonumber \\
&&\left .+{M_Z^2 \over 24 m_{q'}^2} {   a^2 \over   (2+a M_Z)^2} 
-{M_W^2 \over 12 m_{q'}^2} {   a^2 \over   (2+a M_W)^2}   \right]
-{4 \over 3a}\alpha_s +{1\over  a^3} \left[ {a \over m_{q'}}
-{5 \over 4 m_{q'}^3 a} \right]
\\
E[a]^{({\bf 8},0,0)}&=&{2\sqrt{2}G_Fm_{q'}^2 \over \pi a^3}
\left[ -{(1+\delta)a^2 \over   2(2+a m_h)^2} +{1 \over 4m_{q'} ^2} 
+ {m_h^2 \over 8 m_{q'}^2 }{   a^2 \over   (2+a m_h)^2}\right. 
\nonumber \\
&&\left .-{M_Z^2 \over 8 m_{q'}^2} {   a^2 \over   (2+a M_Z)^2} 
-{M_W^2 \over 4 m_{q'}^2} {   a^2 \over   (2+a M_W)^2}   \right]
+{1 \over 6a}\alpha_s +{1\over  a^3} \left[ {a \over m_{q'}}
-{5 \over 4 m_{q'}^3 a} \right]
\\
E[a]^{({\bf 8},0,1)}&=&{2\sqrt{2}G_Fm_{q'}^2 \over \pi a^3}
\left[ -{(1+\delta)a^2 \over   2(2+a m_h)^2}-{1 \over 4m_{q'} ^2}  
+ {m_h^2 \over 8 m_{q'}^2 }{   a^2 \over   (2+a m_h)^2} \right.
 \nonumber \\
&&\left .+{M_Z^2 \over 24 m_{q'}^2} {   a^2 \over   (2+a M_Z)^2} 
+{M_W^2 \over 12 m_{q'}^2} {   a^2 \over   (2+a M_W)^2}   \right]
+{1 \over 6a}\alpha_s +{1\over  a^3} \left[ {a \over m_{q'}}
-{5 \over 4 m_{q'}^3 a} \right]
\\
E[a]^{({\bf 8},1,0)}&=&{2\sqrt{2}G_Fm_{q'}^2 \over \pi a^3}
\left[ -{(1+\delta)a^2 \over  2 (2+a m_h)^2}-{1 \over 4m_{q'} ^2}  
+ {m_h^2 \over 8 m_{q'}^2 }{   a^2 \over   (2+a m_h)^2}\right. 
\nonumber \\
&&\left .-{M_Z^2 \over 8 m_{q'}^2} {   a^2 \over   (2+a M_Z)^2} 
+{M_W^2 \over 4 m_{q'}^2} {   a^2 \over   (2+a M_W)^2}   \right]
+{1 \over 6a}\alpha_s +{1\over  a^3} \left[ {a \over m_{q'}}
-{5 \over 4 m_{q'}^3 a} \right]
\\
E[a]^{({\bf 8},1,1)}&=&{2\sqrt{2}G_Fm_{q'}^2 \over \pi a^3}
\left[ -{(1+\delta)a^2 \over  2 (2+a m_h)^2} -{1 \over 12m_{q'} ^2}  
+ {m_h^2 \over 8 m_{q'}^2 }{   a^2 \over   (2+a m_h)^2} \right. 
\nonumber \\
&&\left .+{M_Z^2 \over 24 m_{q'}^2} {   a^2 \over   (2+a M_Z)^2} 
-{M_W^2 \over 12 m_{q'}^2} {   a^2 \over   (2+a M_W)^2}   \right]
+{1 \over 6a}\alpha_s +{1\over  a^3} \left[ {a \over m_{q'}}
-{5 \over 4 m_{q'}^3 a} \right]
\end{eqnarray}
\end{widetext}
\vskip0.25in


\end{document}